\documentclass[twocolumn,prl,showpacs,preprintnumbers,amsmath,amssymb]{revtex4}
\usepackage{amsmath,amssymb}
\usepackage{graphicx}
\date{\today}
\begin{document}

\title{Magnetic field-induced spectroscopy of forbidden optical
transitions with application to lattice-based optical atomic clocks}

\author{A. V. Taichenachev}
\author{V. I. Yudin}
\affiliation{Institute of Laser Physics SB RAS, Novosibirsk 630090,
Russia}
\affiliation{Novosibirsk State University, Novosibirsk
630090, Russia}
\author{C. W. Oates}
\author{C. W. Hoyt}
\author{Z. W. Barber}
\altaffiliation{also at Univ. of Colorado, Boulder, CO 80309, USA.}
\author{L. Hollberg}
 \affiliation{National Institute of
Standards and Technology, Boulder, CO 80305}

\thanks{Official contribution of the National Institute of Standards and Technology; not subject to copyright.}

\begin{abstract}
We develop a method of spectroscopy that uses a weak static magnetic
field to enable direct optical excitation of forbidden
electric-dipole transitions that are otherwise prohibitively weak.
The power of this scheme is demonstrated using the important
application of optical atomic clocks based on neutral atoms confined
to an optical lattice. The simple experimental implementation of
this method -- a single clock laser combined with a DC magnetic
field-- relaxes stringent requirements in current lattice-based
clocks (\textit{e.g.}, magnetic field shielding and light
polarization), and could therefore expedite the realization of the
extraordinary performance level predicted for these clocks. We
estimate that a clock using alkaline earth-like atoms such as Yb
could achieve a fractional frequency uncertainty of well below
10$^{-17}$ for the metrologically preferred even isotopes.
\end{abstract}


\maketitle

The long coherence times provided by ``forbidden'' excitation to long-lived
atomic states are critical in several important
applications including quantum computing and optical atomic clocks
\cite{Langer05,Diddams04}. However, the search for narrower lines
can lead to transitions between states with undesirably complex
structure, such as levels with relatively large angular momentum. An
important example of an ultra-narrow transition is the
\itshape $^1$S$_0$$\to$$^3$P$_0$ \upshape spin- and angular momentum-forbidden
clock transition in alkaline earth-like atoms. Two of these atoms, Sr and
Yb, are being pursued as strong candidates for
lattice-based optical atomic clocks
\cite{Takamoto05,Courtillot03,Ludlow05,Hoyt05,Hong05b,Park03}. The
long interaction times, high signal-to-noise ratio, and vanishing
Doppler-related uncertainties provided by tight confinement of an
atomic ensemble to individual optical lattice sites should lead to
exceptional short-term stability and high accuracy in this
rapidly developing field.

Current experimental work on the optical lattice clock~\cite{kat02}
focuses on the \itshape $^1$S$_0$$\to$$^3$P$_0$ \upshape transition
in the odd isotopes of Sr and
Yb~\cite{Takamoto05,Ludlow05,Hong05b,Hoyt05}, which is weakly
allowed due to hyperfine mixing. In comparison with their even
counterparts that have zero nuclear spin, however, the odd isotopes
have an uncomfortably large sensitivity to magnetic fields (MHz/T)
and lattice light polarization. They also have multiple ground state
sub-levels that considerably complicate spectroscopic lineshapes.
The method described in this Letter uses a small magnetic field
($\sim$1 mT) to do in a controllable way for the even
isotopes what the nuclear magnetic moment 
does in the odd isotopes: mix a small fraction of the nearby
\itshape $^3$P$_1$ \upshape state into the \itshape $^3$P$_0$
\upshape state, allowing single-photon excitation of the \itshape
$^1$S$_0$$\to$$^3$P$_0$ \upshape clock transition. Estimates for Yb
using experimentally realistic linewidths (0.5 Hz) show that induced
frequency shifts can be controlled at a level that could enable a
fractional frequency uncertainty of well below $10^{-17}$ for a
lattice clock. In contrast to multiphoton methods proposed for the
even isotopes \cite{Santra05,Hong05}, this method of direct
excitation requires only a single probe laser and no
frequency-mixing schemes, so it can be readily implemented in
existing lattice clock experiments. Our method is equally effective
for all alkaline earth-like clock candidates (Yb, Sr, Ca, and Mg).

\begin{figure}[htb]
\centerline{\includegraphics[width=5.5cm]{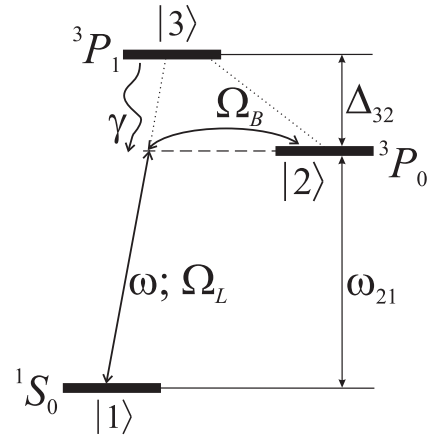}}
 \caption{Magnetic field-induced excitation of a strongly forbidden transition in a generic
three-level atomic system.  A small magnetic
field ($\sim$ mT) mixes excited states $|2\rangle$ and $|3\rangle$, allowing
single-photon excitation (for $\omega = \omega_{21}$) of the otherwise forbidden transition between
states $|1\rangle$ and $|2\rangle$. This approach can work for a number of interesting
alkaline earth-like elements (Yb, Sr, Ca, and Mg), for which the relevant states are labelled
 \itshape $^{2S+1}$L$_J$ \upshape.}
 \label{genericlevels}
\end{figure}


To excite the strongly forbidden $|1\rangle$$\to$$|2\rangle$ clock transition at
frequency $\omega_{21}$ as shown in Fig.~\ref{genericlevels}, we apply a static magnetic field, {\bf B}, that couples
the states $|2\rangle$ and $|3\rangle$.  These states are split by a
 frequency $\Delta$$_{32}$,  and have a
coupling matrix element $\Omega_B$=$\langle 2|$$\hat{\bf \mu} \cdot
{\bf B}$$|3\rangle$/$\hbar$, where $\hat{\bf \mu}$ is the magnetic-dipole
operator. According to first-order perturbation theory with
the condition $|\Omega_{B}/\Delta_{32}|\ll1$, the state $|2\rangle$
acquires a small admixture of the state $|3\rangle$ due to the
presence of the static magnetic field:
\begin{equation}\label{2'}
|2'\rangle=|2\rangle +\frac{\Omega_B}{\Delta_{32}}|3\rangle\,.
\end{equation}
The result is that the transition $|1\rangle$$\to$$|2'\rangle$
becomes partially allowed (\textit{e.g.}, for spontaneous emission
\cite{Ovsyannikov01,Schef05}). An optical field with amplitude
\textbf{E} and frequency $\omega$ (i.e., ${\bf E}(t)={\bf E}e^{-i
\omega t}+ c.c.$) acts via the $|1\rangle$$\to$$|3\rangle$
electric-dipole transition. We assume that this transition is at
least weakly allowed (\textit{e.g.}, an intercombination transition)
and has a decay rate $\gamma$. The corresponding coupling matrix
element is the Rabi frequency, $\Omega_L$=$\langle 3|$$\hat{{\bf
d}}\cdot {\bf E}$$|1\rangle$/$\hbar$, where $\hat{\bf{d}}$ is the
electric-dipole operator. Due to the slight admixture of state
$|3\rangle$ into the bare state $|2\rangle$, a resonance will be
observed on the forbidden transition ($|1\rangle$$\to$$|2\rangle$)
when $\omega \approx \omega_{21}$. Using Eq.~\ref{2'}, the induced
transition rate is
\begin{equation}
V_{12}=\langle
2'|\hat{{\bf d}}\cdot {\bf E}|1\rangle/\hbar =\frac{\Omega_L\,\Omega_B}{\Delta_{32}}\,. \label{V12}
\end{equation}
Remarkably, we find that a reasonable excitation
 rate ($\sim$1~Hz) results from very modest field values (\textit{e.g.}, B $\sim$1~mT and light
intensity $\sim$10~mW/cm$^2$) for realistic atomic parameters.

This result can also be interpreted from the perspective of
two-photon spectroscopy. In this case the expression for V$_{12}$ in
Eq.\ \ref{V12} is the on-resonance two-photon Rabi frequency.
Specifically, if the frequency of the field connecting $|2\rangle$
to $|3\rangle$ is set to zero, the condition for the usual
two-photon resonance (i.e., zero two-photon detuning from transition
$|1\rangle$$\to$$|2\rangle$) becomes $\omega = \omega_{21}$. The
frequency splitting $\Delta_{32}$ plays the role of one-photon
detuning.

Due to the small admixture of $|3\rangle$, state $|2'\rangle$ has a
finite lifetime, which leads to broadening of the forbidden
transition.  Including power broadening due to the laser light, the
total broadening can be estimated in the framework of perturbation
theory as
\begin{equation}\label{g12}
\gamma_{12}\approx \gamma \frac{\Omega^2_L+\Omega^2_B}{\Delta^
2_{32}}.
\end{equation}
This broadening effectively comes from the fraction of population
transferred by the two fields to the $|3\rangle$ state, which decays
with a rate $\gamma$.  Since this broadening is inversely
proportional to the square of the large frequency splitting
$\Delta_{32}$, it is an extremely small quantity.  For typical field
values, the resultant broadening will be much less than 1~$\mu$Hz.

Of considerable importance to clock applications are the quadratic
shifts of the transition frequency $|1\rangle$$\to$$|2\rangle$ that
result from the applied fields.  For a simple three-state system,
the optical Stark shift is
\begin{equation}\label{L_sh}
\Delta_L=\frac{\Omega^2_L}{\Delta_{32}}\,,
\end{equation}
while the second-order Zeeman shift is
\begin{equation}\label{B_sh}
\Delta_B=-\frac{\Omega^2_B}{\Delta_{32}}\,.
\end{equation}
The excitation rate in Eq.\ \ref{V12} can then be re-written in terms
of the induced field shifts,
\begin{equation}\label{V12_sh}
V_{12}=\sqrt{|\Delta_L\Delta_B|}\,.
\end{equation}
In the example below we will see that reasonable excitation rates
lead to shifts of a few hertz or less, which can be controlled at a
much finer level. Moreover, the same induced transition rate can be
realized with different combinations of $\Delta_L$ and $\Delta_B$,
which allows a high degree of experimental flexibility. For example,
if the magnetic field is more easily controlled than the optical
field, it could be preferable from the metrological viewpoint to
work under the condition $\Delta_B > \Delta_L$.

This flexibility is in stark contrast to the case of the odd
isotopes.  Due to their intrinsic nuclear magnetic moments, these
atoms have fixed effective magnetic fields ($\sim$ 1 T), which are
much larger than the applied fields we propose to use ($\sim$~1 mT).
The nuclear magnetic field mixes a much larger fraction
($\sim$1000x) of $|3\rangle$ into $|2\rangle$ (see Eq.~\ref{2'}),
producing a natural linewidth of $\sim$ 15 mHz for the forbidden
transition in Yb.  As a result, relatively little probe laser
intensity (a few~$\mu$W/cm$^2$) is required to excite the forbidden
transition.  In exchange, however, one finds a much larger magnetic
sensitivity with a first-order dependence on the B field of several
MHz/T.  For reasonable experimental parameters, the light shifts are
negligible (sub-mHz), but the magnetic shifts are large, requiring
field control (or isolation) at the 100 pT (microgauss) level. One
of the strengths of our method is that we can tune the size of the
magnetic field so that the uncertainties resulting from the induced
shifts are of similar size. As a result, the magnetic field
shielding requirements are relaxed by a factor of $\sim$1000,
whereas the light shifts remain at a manageable level (see example
below), greatly simplifying the experimental apparatus.

Let us now expand our discussion based on a three-level atom to
account for the level structure of a real atom (refer again to
Fig.~\ref{genericlevels}), using the example of even isotopes of
alkaline earth-like atoms (\textit{e.g.}, Mg, Ca, Sr, and Yb). Our
objective is to excite the forbidden (but tantalizing) \itshape
$^1$S$_0$$\to$$^3$P$_0$ \upshape optical clock transition, using the
intercombination transition \itshape $^1$S$_0$$\to$$^3$P$_1$
\upshape to supply the required electric-dipole interaction. The
magnetic-dipole coupling with the static field {\bf B} is
implemented via the \itshape $^3$P$_1$$\to$$^3$P$_0$ \upshape
transition. We note that the lowest-lying \itshape $^1$P$_1$
\upshape state contributes to the induced transition rate in the
same way as the \itshape $^3$P$_1$ \upshape state, but at a reduced
level due to its approximately ten times larger detuning.

Taking into account the vector nature of the applied fields and the
Zeeman degeneracy of the level \itshape $^3$P$_1$ \upshape , the
expression for the induced transition rate in Eq.\ \ref{V12} takes
the specific form
\begin{equation}\label{V_EB}
V_{12}=\frac{\langle \| d \| \rangle\,\langle \| \mu \|
\rangle ({\bf E \cdot B})}{\hbar^2\Delta_{32}}\,.
\end{equation}
Here $\langle \| d \| \rangle$ is the reduced matrix element of the
electric-dipole moment on the \itshape $^1$S$_0$$\to$$^3$P$_1$ \upshape transition,
and $\langle \| \mu \| \rangle$ is the reduced matrix element of the
magnetic-dipole moment on the \itshape $^3$P$_1$$\to$$^3$P$_0$ \upshape transition.
For even isotopes of the alkaline earth-like elements, $\langle \|
\mu \| \rangle = \sqrt{2/3}~\mu_B$, where $\mu_{B}$ is the Bohr
magneton.  Values for the fine-structure splittings ($\Delta_{32}$)
and electric-dipole matrix elements, however, are strongly
atom-dependent (see Table 1). To evaluate the applicability of this method for
various atoms, it is useful to rewrite the expressions for the excitation rate (V$_{12}$)
and the induced shifts ($\Delta$$_{L,B}$) in terms of the applied
fields.  Combining all the constant terms into a single value, $\alpha$, the
expression in Eq.\ \ref{V_EB} can be re-written
\begin{equation}\label{V_IB}
V_{12} = \alpha \sqrt{I}\,|{\bf B}|\cos \theta\,,
\end{equation}
where $\alpha$ is a measure of the induced transition rate per unit
of each of the fields, $I$ is the light field intensity, and
$\theta$ is the angle between linearly-polarized \textbf{E} and \textbf{B} fields.
Similarly, the quadratic Zeeman shift can be written
\begin{equation}\label{shB_Yb}
\Delta_B=\beta |{\bf B}|^2\,,
\end{equation}
while the light shift
on the transition frequency $\omega_{21}$ can be expressed as
\begin{equation}\label{shE_Yb}
\Delta_L=\kappa I\,,
\end{equation}
where $\beta$ and $\kappa$ are the respective shift coefficients.
For correct estimates of the light shift (i.e. the coefficient
$\kappa$ in Eq.\ \ref{shE_Yb}) it is necessary to take into account
the contributions of all states connected by dipole transitions to
working levels \itshape $^1$S$_0$ \upshape and \itshape $^3$P$_0$ \upshape . In particular, the main
contributions to the shifts on \itshape $^1$S$_0$ \upshape and \itshape $^3$P$_0$ \upshape are the
nearest respective \itshape $^1$P$_1$ \upshape and \itshape $^3$S$_1$ \upshape states. Using
 Eqns.~\ref{shB_Yb} and \ref{shE_Yb}, Eq.~\ref{V_IB} can be written
\begin{equation}\label{V12_shYb}
V_{12}=\xi\sqrt{|\Delta_L\Delta_B|}\,\cos \theta\,,
\end{equation}
where $\xi$ $\equiv$ $\alpha$/$\sqrt{\beta \kappa}$. The factor
$\xi$ can be considered a dimensionless ``quality'' factor for the
clock transition in this scheme, since it relates the strength of
the excitation to the magnitudes of the induced field shifts. A
larger value for $\xi$ implies that for a given spectroscopic
linewidth, the induced shifts will be smaller.  Note that $\xi=1$
when only the light shift on the ground state due to level \itshape
$^3$P$_1$ \upshape is considered, as in a simple three-level case.

Table \ref{table} summarizes the relevant parameters for four
different alkaline earth-like atoms that have been considered as
excellent clock candidates \cite{Porsev04, kat02, rus98, ste04a}.
Despite the wide range of intercombination transition line strengths
(given in Table 1 as natural decay rate $\gamma$) and fine structure
splittings ($\Delta$$_{32}$), we find similar values for $\alpha$
and $\xi$.  Therefore the scheme proposed here may be applied to any
of these atoms. This opens the door to new clock possibilities,
especially for Mg and Ca, for which multi-photon schemes designed to
use the even isotopes require experimentally inconvenient lasers
(\textit{e.g.}, in the visible and IR range not covered by laser
diodes).  We do emphasize, however, that the level shifts
($\Delta_{L,B}$) will generally be different for the same light
intensities and magnetic field magnitudes. Thus, different atoms
will generally require different combinations of field values to
achieve comparable induced transition rates ($V_{12}$) while keeping
field shifts manageable. The values for $\kappa$ in Table 1 have an
uncertainty of a few mHz/(mW/cm$^2$) due to uncertainties in
transition rates to higher lying states, leading to corresponding
uncertainties in $\xi$.

\begin{table}[ht]
\caption{Atomic species comparison}
\begin{tabular}[t]{| c | c | c | c | c | c | c |}
\hline  &  $\gamma/2\pi$  & $\Delta_{32}/2\pi$  & $\alpha$
 & $\beta$
  & $\kappa$  & $\xi$ \\[1pt]
\hline & kHz & THz & $\frac{\mbox{Hz}}{\mbox{T}
 \sqrt{\mbox{mW/cm}^{2}}}$
 &$\frac{\mbox{MHz}}{\mbox{T$^{2}$}}$&$\frac{\mbox{mHz}}{\mbox{mW/cm}^{2}}$  & \\[1pt]
\hline\hline
Yb   &  182    &   21 & 94  &    $-$6.2  & 15    & 0.30 \\
Sr   &       7     &   5.6   & 99  &  $-$23.3 &   $-$18    & 0.15 \\
Ca   &       0.4   &   1.5   & 77  & $-$83.5&   $-$3.5  & 0.14 \\
Mg   &        0.07    &  0.6    & 47  &   $-$217  &   $-$0.5   & 0.14  \\
 \hline
\end{tabular}
\label{table}
\end{table}

Using the values for Yb, let us estimate the frequency uncertainty
induced by the fields for our method, using realistic field
magnitudes and assumptions about their control. A magnetic field
$|{\bf B}|$=1~mT (10~G) with uncertainty $\sim 10^{-7}$~T (1~mG)
leads to a quadratic Zeeman shift of $\Delta_B$$\approx 2 \pi
\times$ 6.2~Hz with an uncertainty $\delta(\Delta_B)\approx 2 \pi
\times$ 1.2~mHz. With a probe laser intensity $I$= 8~mW/cm$^{2}$, we
estimate the light shift to be $\Delta_L \approx 2 \pi \times$
120~mHz, a value that can be verified experimentally. If we assume
that the intensity can be controlled on the level of 1~\%, then the
resulting light shift uncertainty is $\delta(\Delta_L)\sim 2 \pi
\times$ 1.2~mHz. In this case the combined frequency uncertainty due
to these two effects is estimated to be $2 \pi \times$ 1.7~mHz,
corresponding to a fractional frequency uncertainty of $3 \times
10^{-18}$ on the forbidden transition frequency
$\omega_{12}=2\pi\times$ 518~THz.

For these values of the fields the induced transition rate given in Eq.~\ref{V_IB} with ${\bf E}||{\bf B}$ is
$V_{12}$$\approx 2 \pi \times$ 0.26~Hz, a
 convenient value for optical atomic clock studies.
This induced Rabi frequency would yield a Fourier
transform-limited feature with an approximate width of 0.5~Hz.
Resolution of the peak would require a laser linewidth that is within present laser stabilization capabilities
\cite{Young99}. Due to its lower sensitivity to field shifts in comparison
with proposed multi-photon schemes for the even isotopes \cite{Santra05,Hong05}, this method enables
 a more experimentally-accessible excitation rate while keeping
the field shifts well below 10$^{-17}$. If we set our linewidth
 to the more experimentally challenging value of $\sim$10 mHz (as assumed in Refs.\
\cite{Santra05,Hong05}), our field shift uncertainties ($\delta(\Delta_{L,B})$) contribute
 less than 1 part in than 10$^{19}$. Of course, other
potential shifts need to be considered when determining the total
uncertainty for an absolute frequency measurement. For example, a
particular concern for a high-precision Yb clock is the shift due to
blackbody radiation, which is expected to be approximately
$-$1.25~Hz at room temperature with a temperature dependence of
$\sim$17~mHz/K \cite{Derevianko05}. 
The evaluation of such small shifts will benefit greatly from the
high stability that could be achieved with a lattice-based system.
With a sample of only $10^{4}$ atoms, a single measurement with the
required 2~s interaction time could yield an imprecision approaching
$10^{-17}$, assuming adequate laser pre-stabilization.

The necessary interaction time can be provided by a non-dissipative
optical lattice, which holds the atoms against gravity. The lattice
wavelength is chosen to yield equal Stark shifts for the upper and
lower clock states ($|1\rangle$ and $|2\rangle$ in
Fig.~\ref{genericlevels}) \cite{kat02,Takamoto05,Ludlow05}. 
The lattice-induced decoherence time is expected to be small
compared to our projected interaction time of 2~s. An optical clock
based on odd-isotope Yb atoms confined to such a shift-cancelling
lattice has been projected by Porsev \textit{et al.} \cite{Porsev04}
to have a potential fractional frequency uncertainty of $10^{-17}$.
With the scheme proposed here we could potentially achieve an even
lower uncertainty, since the even isotopes have less magnetic field
sensitivity and no complications due to the multiplicity of the
ground state (\textit{e.g.}, optical pumping and lattice
polarization sensitivity). Furthermore, spectroscopy based on a
single ground state will simplify the determination of the
unperturbed transition frequency since only a single spectroscopic
feature will be present (instead of a multi-peaked Zeeman spectrum).

A particularly attractive aspect of this approach is that it could
be implemented in current odd isotope lattice clock experiments with
minimal change to the apparatus. In contrast to the other schemes
for the even isotopes \cite{Santra05,Hong05}, our method requires
neither additional probe lasers nor associated non-linear optics to
perform requisite sum/difference frequency generation.  Instead only
a modest magnetic field needs to be generated, which can be done
with a pair of current-carrying Helmholtz coils.

In summary, we have shown that strongly forbidden transitions can be
accessed via single-photon excitation with the application of a
relatively weak static magnetic field.  With this method reasonable
excitation rates can be achieved without inducing significant Stark
shifts, Zeeman shifts, or line broadening. Using Yb as an example,
we have demonstrated how this approach could enable an optical
lattice-based clock using the even isotopes of alkaline earth-like
atoms. The even isotopes are far less sensitive to magnetic fields
and have much simpler structure than their odd counterparts. We also
note that this method can be adapted to enable transitions between
states $|1\rangle$ and $|2\rangle$ of the same parity (but with
state $|3\rangle$ of opposite parity) by replacing
 the static magnetic field with a static electric field. 

This work was supported by RFBR (grants 05-02-17086, 04-02-16488,
and 05-08-01389) and by a grant INTAS-01-0855


%
%

\end{document}